\documentclass[onecolumn]{rsauthor} 
\usepackage{graphicx} 

\jname{Phil. Trans. R. Soc. A}

\artnum{XXXXXXX}

\begin{document} 

\title
{Interpreting signals from astrophysical transient experiments}

\author[P.T. O'Brien, S. Smartt]{Paul T. O'Brien$^1$, Stephen J. Smartt$^2$} 

\address{$^1$Department of Physics and Astronomy, University of Leicester,  
University Road, Leicester, LE1 7RH, UK \\
$^2$Astrophysics Research Centre, School of Mathematics and Physics,
Queen's University Belfast, Belfast, B17, 1NL, UK}

\date{...}

\abstract{
Time domain astronomy has come of age with astronomers now able to
monitor the sky at high cadence both across the electromagnetic
spectrum and using neutrinos and gravitational waves. The advent of
new observing facilities permits new science, but the ever increasing
throughput of facilities demands efficient communication of coincident
detections and better subsequent coordination among the scientific
community so as to turn detections into scientific discoveries. To
discuss the revolution occurring in our ability to monitor the
Universe and the challenges it brings, on 2012 April 25-26 a group of
scientists from observational and theoretical teams studying
transients met with representatives of the major international
transient observing facilities at the Kavli Royal Society
International Centre, UK. This immediately followed the Royal Society 
Discussion meeting  ``New windows on transients across the
Universe'' held in
London.   Here we present a summary of the Kavli meeting at
which the participants discussed the science goals common to the
transient astronomy community and analysed how to better meet the
challenges ahead as ever more powerful observational facilities come
on stream.}



\maketitle 

\section{Introduction} 

Astronomy has always progressed by taking advantage of the latest
observing techniques and combining information from across the
electromagnetic spectrum. This approach has led to the modern subject
of astrophysics in which sophisticated models have been constructed to
explain everything from planets to the origin of the Universe. These
models are largely based on either a static or slowly evolving set of
data. A revolution is now beginning in terms of our ability to monitor
the Universe using both existing and new observing facilities, such
as the LOFAR radio telescope, the IceCube neutrino observatory, and
a range of optical telescopes with of order 10 square degree imaging
capabilities (Pan-STARRS1, Palomar Transient Factory, La Silla-QUEST, 
SkyMapper, Catalina Real Time Survey).   The data from these ground-based
facilities when combined with those from orbiting high-energy
observatories, such as Swift, Fermi and MAXI, allow us to say that
multi-messenger astronomy has now come of age. 
Gravitational wave experiments (LIGO, VIRGO) have been running now 
for some years. While detection still awaits, the upgrade path to
Advanced LIGO promises exciting opportunities in opening a new window
in astrophysics.  In this new paradigm,
astronomers are aiming to capture the behaviour of the temporal sky with
high cadence across the electromagnetic spectrum and through 
neutrino and gravitational wave physics. 

Using the temporal domain enables a census of the transient sky and
permits detailed study of individual objects which test our
understanding of the laws of physics under the most extreme
conditions. The best known examples of luminous extragalactic
transients involve the death of massive stars or the merger of compact
objects which produce phenomena such as supernovae and Gamma-Ray
Burst  (Gehrels 2013, Hjorth 2013, Piran 2013). 
Numerous types of less luminous but astronomically important
transients involving stellar flaring or accretion processes are also
seen in our Galaxy. Overall, the importance of the temporal domain in
terms of understanding the evolution of stars and galaxies and how
they interrelate has been recognised in many recent reports, such as
those by the European Union ASTRONET group and the USA National
Research Council Decadal Survey.

The increasing rate of transient detection and observation cadence
provides an opportunity to accurately track transient evolution and
can bring relatively rare events into view which can greatly increase
our understanding. But with the opportunity also comes a threat that
our previous ways of working and communicating may be inadequate to
cope with the data deluge. On 2012 April 25-26 a group of about 
50\footnote{List of participants :
	Stephane	Basa,	
	John	Beacom,	
	Lars	Bildsten,	
	Joshua	Bloom,	
	David	Burrows,	
	Sergio	Campana,	
	Valerie	Connaughton,	
	Bertrand	Cordier,	
	Paul	Crowther,	
	Melvyn	Davies,	
	Rob	Fender,	
	Andrew	Fruchter,	
	Neil	Gehrels,	
	Jochen	Greiner,	
	Ik Siong	Heng,	
	Jim	Hinton,	
	Simon	Hodgkin,	
	Isobel	Hook,	
	Susumu	Inoue,	
	Nobuyuki	Kawai,	
	Andrew	Levan,	
	Andrew	MacFadyen,	
	Sera	Markoff,	
	Julie	McEnery,	
	Carole	Mundell,	
	Paul	O'Brien,	
	Julian	Osborne,	
	Stephan	Rosswog,	
	Sheila	Rowan,	
	Norbert	Schartel,	
	Brian	Schmidt,	
	Stephen	Smartt,	
	Benjamin	Stappers,	
	Patrick	Sutton,	
	John	Swinbank,	
	Gianpiero	Tagliaferri,	
	Nial	Tanvir,	
	John	Tonry,	
	Alexander	van der Horst,	
	Ralph	Wijers,
	Alicia 	Soderberg,	
	Jianyan	Wei,	
	Andy 	Lawrence,	
	Matt	Page,	
	Silvia 	Zane,	
	Jean-Luc 	Atteia} 
astronomers working in the area of transient astronomy met under the
auspices of The Royal Society to discuss their common science goals
and ways to enhance the use of multi-messenger facilities. 
This gathering followed immediately after the Royal 
Society Discussion Meeting in London on ``New windows on transients across the
Universe'', the papers of which are presented in this edition
\footnote{Talk slides are available here
  http://www.star.le.ac.uk/$\sim$pto/RS2012/}.  The
participants were divided into three groups (led by Lars Bildsten,
Neil Gehrels and Brian Schmidt) each charged with the same tasks to
see if any consensus could be achieved and to inject a note of
competition. This paper summarises the outcome on the topics of: (1)
what are the key science questions?, (2) how do we better coordinate
the use of facilities? and (3) what future facilities are envisaged?

\section{Science drivers}

\begin{figure}
\begin{center}
 \includegraphics[angle=90, width=0.7\textwidth]{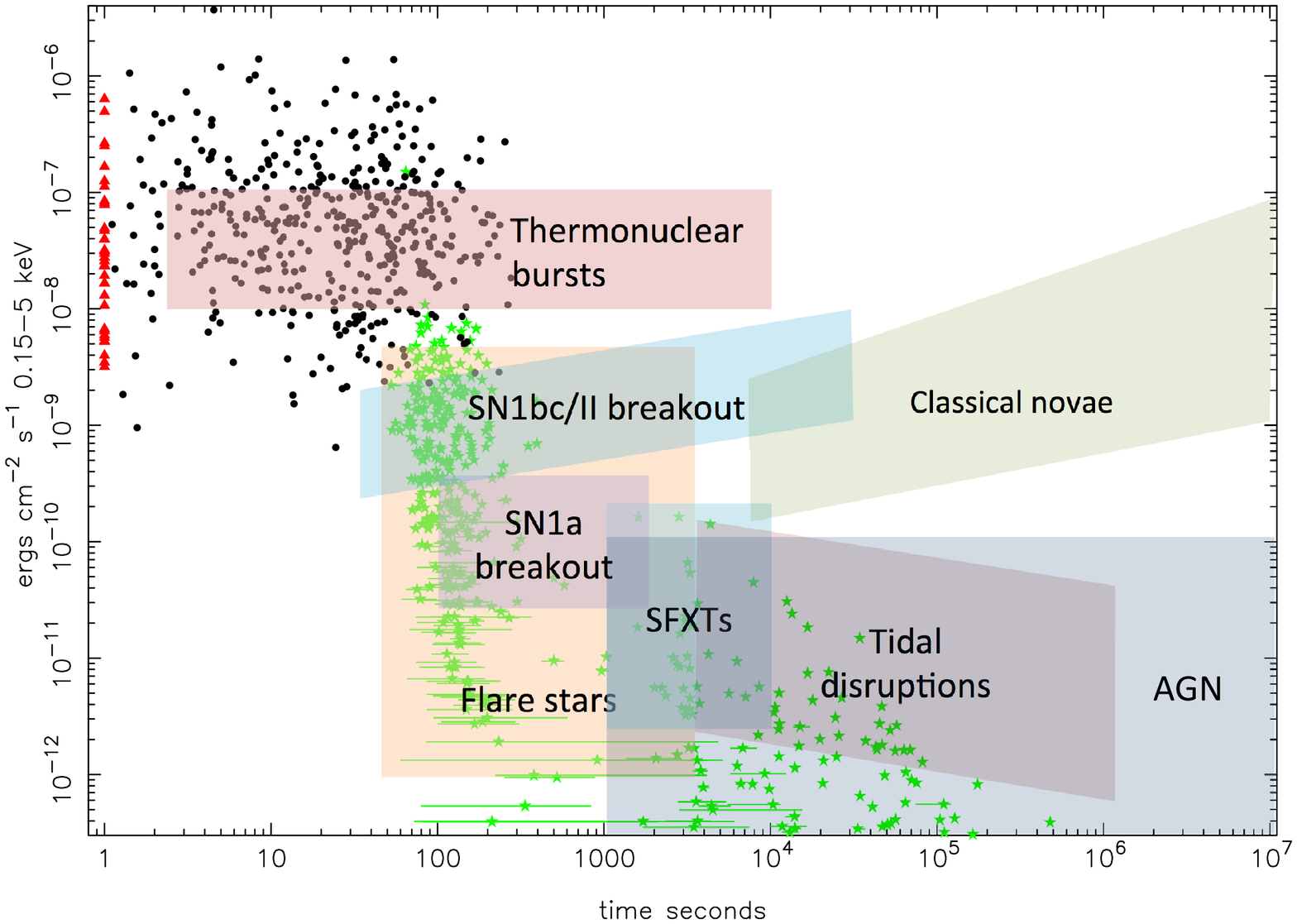}
\includegraphics[angle=-90, width=0.7\textwidth]{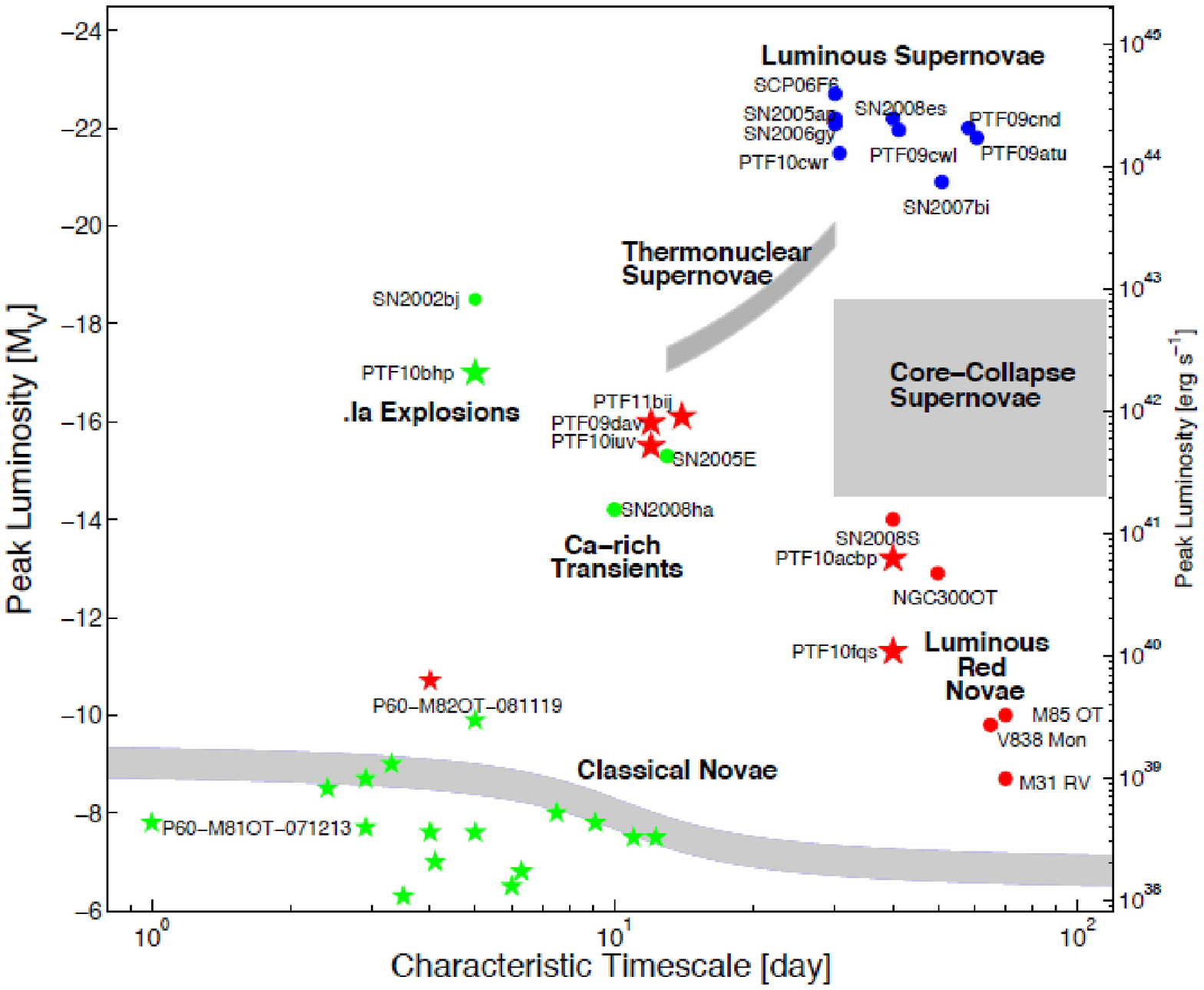}
\end{center}
 \caption{{\bf Upper panel:} shows the typical observed X-ray flux
 plotted against variability timescale for a variety of source types
 (colour shaded regions) and for the prompt and afterglow fluxes for
 GRBs  detected by the Swift mission (individual points). 
Black points are Swift BAT GRBs (with the $T_{90}< 1$ sec in
red), green points are Swift XRT GRB afterglow fluxes. 
{\bf Lower panel :}  taken from Kulkarni (2013) and shows the optical phase space of
 cosmic explosive events and their characteristic timescales. Image
 Credits : Julian Osborne (upper) and Shri Kulkarni (lower). }
\label{figure1} 
\end{figure}

Traditionally, wide-field sky monitoring has been confined either to
observing high-energies from space (e.g. the detection of gamma-ray
bursts, GRBs) or to the optical (e.g. the detection of novae and
supernovae). While existing gamma-ray detectors can survey large sky
areas (thousands of square degrees) simultaneously on milli-second
timescales, optical surveys have tended to monitor small sky areas
(few to a few tens of square degrees) with a cadence from several
hours to days or weeks (Tonry 2013). This observation strategy has differentiated
the science discoveries at the shortest timescales but there is
considerable overlap in the source types monitored on longer
timescales. Some of the transient source types monitored in
optical and X-ray astronomy are shown in Figure~\ref{figure1}.

\begin{figure}
\begin{center}
 \includegraphics[angle=-90, width=0.5\textwidth]{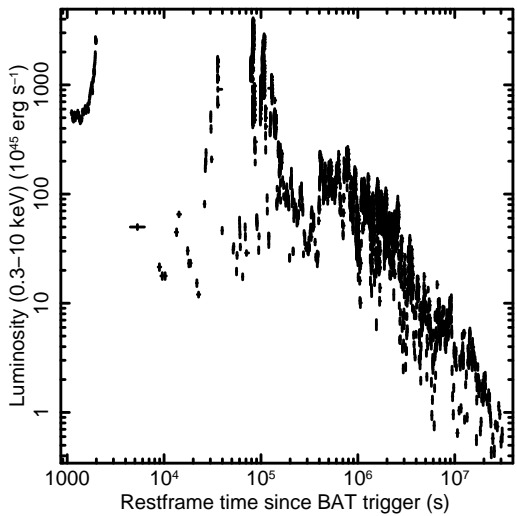}
\end{center}
 \caption{The late time 0.3-10 keV X-ray light curve of Swift
 J164449.3+573451, a
 relativistic tidal disruption event at z=0.3543 (Bloom et al. 2011;
 Levan et al. 2011; Burrows et al. 2011). The data were taken from the
 UK Swift data centre site 
\protect\url{http://www.swift.ac.uk/xrt_curves/00450158/}
(Evans et al. 2009). Even allowing for modest beaming this event
has a luminosity after a year which is comparable to that of a bright
AGN.}
\label{figure2} 
\end{figure}

The more recently commissioned monitoring facilities are starting to
fill in sections of the phase space shown in Figure~\ref{figure1}. For
example :  PTF,  PS1 and La Silla-Quest have detected several
luminous supernovae while the Swift satellite and the PS1+GALEX
combination  have detected several
long-lived transients which appear to be tidal disruption events (TDEs)
(e.g. Figure~\ref{figure2}). Once detected these sources have to be
classified by bringing together information from multiple follow-up
facilities. 

The three breakout groups all came up with broadly similar views of
the  high priority outstanding science questions to address in the broad field of 
astrophysical transients.

\begin{enumerate}
\item Extreme physics --- the study of black holes, neutron stars, compact
binary evolution, supernovae, GRBs and tidal disruption events. 
\begin{enumerate}
\item What are the explosion mechanisms for massive stars and how many 
progenitor types are there that produce the observed diversity ? 
\item What exactly are the progenitors of thermonuclear supernovae and
 how many channels can produce explosions ?  
\item  Can we confirm that compact binary mergers are the origins of
 short GRBs --- can coincident photonic and gravitational  wave
 detection be realised for these objects ?
\item What is the relative importance of accretion and jet power in 
AGN, GRBs and tidal disruption events and what makes the
$>10^{18}$\,eV cosmic rays ?
\item Can we determine the equation of state of neutron stars using gravitational 
waves, neutrinos and electromagnetic timing data.
\item Are there ultimate physical limits in energy and time to
 explosive events ? 
\end{enumerate}

\item Gravity beyond Einstein --- testing our understanding of 
relativity and determine the properties of dark matter and dark energy.
\begin{enumerate}
\item Test the propagation and polarisation of gravitational waves.
\item Test Lorentz invariance using light from radio to gamma-ray energies. 
\end{enumerate}

\item Transients as probes --- use transients as a means to constrain cosmology
and illuminate the early universe.
\begin{enumerate}
\item Determine the nature of dark energy using cosmic probes, primarily 
via supernovae but possibly also gravitational wave sirens.
\item Use GRBs as bright beacons to locate the first stars and  galaxies.
\end{enumerate}
\end{enumerate}

The meeting attempted to distill the over arching science questions 
into ideas for future coordination, building of facilities and future
missions, and working coherently together.  While this diverse group
obviously had many opinions on the exciting science areas that could
be opened up in this regime, two  particular areas stood out which
could provide  potentially ground-braking results if facilities can
either work together or new, affordable, missions are built. These
were the first gravitational wave sources, which are most likely to be 
NS-NS or NS-BH mergers, and ensuring that GRBs can be used to probe
the  high redshift Universe and the first stars. In Section\,3 we
describe what 
the meeting participants thought  about current facilities and how
they could work together more coherently. In Section\,4
we discuss  the facilities required to target these areas, and how the 
science goals could be realised.

\section{Opportunities and threats}
The exciting science opportunities from the combinations of multi-wavelength and
non-photonic experiments is potentially threatened by how well we do
(or don't) design communication channels. The synergy and direct link
between experiments is effectively another requirement that we should
build into the design of the facilities.

An alert --- a message announcing a new transient event --- can in
principle be issued for many types of transient and
electromagnetic/non-photonic signals. For example, gamma-/X-ray (AGN,
GRBs, TDEs), UV/optical (supernova, TDEs, novae), radio (supernovae,
GRBs, TDEs), neutrino (supernovae, AGN, GRBs) and gravitational waves
(GRBs, AGN). The meaning of an alert varies somewhat among the
transient monitoring facilities, but as a minimum usually involves the
release of a time stamp and a sky position for the transient, which
can be accompanied by a flux/magnitude and a source type
classification. There is a tension between providing rapid alerts with
this basic information and waiting to provide enough information
such that the basic alert is usable to a scientist. For example, one does
not want to be overwhelmed with variable stars when searching for supernovae
in difference images of optical fields.  While that particular problem
has been solved for PTF and PS1 (by simple catalogue cross-matching
where reliable star-galaxy separation exists), a more subtle problem
is finding rare high-redshift transients beyond the foreground fog of
the ubiquitous type Ia SNe. One might be able to declare that a
transient is an extragalactic SN of some sort, even with only 2
reliable detections, but beyond that it is difficult to further
classify without a spectrum or a full light curve. If one waits for the
latter, the interesting early explosion phase is an opportunity
missed. The capability of getting multi-wavelength data at these
physically interesting early stages can be lost if one waits to
confidently classify a particular transient. This tension will almost
certainly increase in future as the rate of transient detections
increases e.g. multi-wavelength triggers from LOFAR, GAIA, HAWC,
ATLAS, and finally LSST (e.g. see Hodgkin 2013, Tonry 2013, Hook 2013). 
 The challenge will often be classifying the
physically different and interesting events early enough that
observations which probe the very earliest stages are ensured without
wasting resources. The latter issue is not just one of wasting
telescope time, humans will often tire of chasing transient events
which have a low probability of being new and interesting.

Transient monitoring facilities are in general operated by dedicated
science consortia. The bulk data are not usually made public rapidly,
and sometimes not made public at all, but alerts usually are on a variety of
timescales.  Examples include PTF, Pan-STARRS-1 Skymapper, La Silla
Quest, IceCube, ANTARES, LOFAR, MAXI, Advanced-LIGO etc.  There are a
few facilities (e.g., Swift) where the data become public after either
no delay or a short delay.  The most rapidly-released truly public
information (i.e. that which anyone can access) tend to come from GRB
facilities as for GRBs rapid follow-up is absolutely essential given
their fast decay rates. Even when data and alerts are made public
rapidly, there are many issues limiting science opportunities such as
how to best disseminate the information, how to ensure good
coordination of subsequent observations and how to provide adequate
scientific reward to those providing the original triggers so as to
maintain motivation and funding. We give some examples of strengths
and weaknesses for some representative transient facilities in
Table~1. There is a common thread to the weaknesses among many facilities and can be
approximated under the headings of communication and coordination. 

\subsection{Communication}
The different communities in transient research 
tend to use different means to communicate, related to the traditions
of that field. 
\begin{enumerate}
\item The GCN (Gamma-Ray Coordinates Network) system run from GSFC
was originally developed for the CGRO-BATSE era to provide alerts for
GRB detection and follow-up observations. GCNs are used by a large
number of facilities and can be delivered via email, text messaging,
web pages, etc. They come in three different flavours. ``Notices'' are usually
automatically generated, provide source locations but can also have
appended small amounts of data such as a light curve. ``Circulars'' are often
hand written and provide a description of data analysis results,
updates on follow-up observations (e.g. a redshift) and can announce
intentions to do something in order to encourage collaboration. 
``Reports'' are hand written and summarise what has happened (currently
only issued by the Swift project).
\item VOEvent, an international system to define the content and 
meaning of standard packets of information about a celestial
event. The system is targeted at automated systems which can generate
and translate VOEvent packets and use them to trigger telescopes,
build web pages, etc. (GCN notices have similar
functionality). VOEvent is used by some facilities (e.g. Swift) often
in parallel to GCNs. Some future facilities, particularly LSST, plan
to use VOEvent for all reporting.
\item ATEL, The Astronomers Telegram, uses HTML formatted hand written text to 
report and comment on observations of transients. They are submitted
via a web interface and are mostly used for `slower' transients, such as
supernovae, but have a wider audience in terms of research fields than
GCNs.
\item IAU circulars, the oldest and most widely recognised means of 
communication is run from Harvard and today is mostly used for
distribution of alerts for reporting on novae, supernovae and comets
(with an analogous system for reporting on minor planets). These
circulars are distributed via a subscription service although some
circulars are freely available. There is a general feeling that this
process is no longer fit for purpose. For example PTF have discovered
and classified $\sim1700$ SNe to date, Pan-STARRS1 has discovered
$\sim$3000 SNe candidates and spectroscopically confirmed
$\sim300$. However neither project has pursued IAU circulars or IAU
names for the transients. The professional discovery industry has now
moved beyond the use of circulars, although they do still server an
important purpose for amateur astronomers to get recognition for their
discoveries and as such it is an essential public outreach tool.
\end{enumerate}

The variety of communication systems used by different parts of
the community discussed above illustrate the disparity between automated
and human-driven alerts and how well-intentioned standards may not be
accepted by the community. Disparities lead to delays and
errors. VOEvent was conceived as a means to solve some of those
issues, and it is relatively well defined under international
agreements. But, it is managed in a way that is slow to respond to
community requests for changes (e.g. the current lack of
timestamps). The management of the GCN system in contrast is quick to
respond to facility/user requests, but it is debatable whether it is
best suited for the multi-messenger era if transients detection rates
increases by many orders of magnitude, as they will with LSST for
example. ATELs and IAU Circulars would certainly be hard to use were
transients discovered at large rates. It is fair to say that there
has, as yet, been no community buy in to a standard way of
communicating across wavelength regimes and each community has tended
to develop what works for them.

Those have tended to be relative successful solutions. But the vision
from this Kavli meeting was that a uniform and standard reporting
method that would facilitate communication on all timescales and
information is both desirable and achievable.  The majority view was
to adopt VOEvent as the standard but with the requirement of changes
in defining and adapting the standards more rapidly to better respond
to the user community.

\subsection{Coordination}

The variety of transient types and follow-up requirements make the
issue of coordination perhaps the greatest challenge in future for
transient research, particularly if transient detection rates
dramatically increase. Increased detection rates not only make
communication harder, but also drive us to use automated,
scientifically well-defined classification schemes. Initial
classification most likely requires some degree of follow-up of a
detection (the PTF integrated approach of discovery and rapid
follow-up with coordinated resources is a recent example), although
multi-wavelength detection coupled with matching of source properties
across on-line, large-area databases can permit some source class
classification automatically.  

Following initial classification (it's a GRB, a supernova, etc.), the
next requirement is to maximise efficient further follow-up. Today
there are numerous examples where coordination is barely
acceptable. These include (a) heterogeneous TOO (target of
Opportunity) procedures among observing facilities even within the
same waveband; (b) no centralised information available on who has triggered
what facility (wasting human and observing time on multiple requests); and
(c) lack of rapid response speed from some facilities due either to
inherent capability (e.g. HST, Chandra) or lack of a clear TOO procedure.

The majority view from the Kavli meeting discussion was that
coordination needs to improve.  Especially in the era of
multi-messenger
signals, for example from high energy transient emission (Hinton
2013) and future radio surveys (Stappers 2013). 
One option would be to have a central
clearing house approach to TOOs. Requests could be submitted to
multiple telescopes and the one best-placed, with clear skies, carries
out the observation. Demanding that all future publicly-funded
facilities are designed for rapid response to TOOs would also help as
would allowing space mission teams to request coordinated ground-based
follow-up as part of the mission proposal. Such changes would
come at the cost of large consortia, perhaps discouraging younger
scientists, but changes to proprietary periods and data rights
could be put in place so as to enhance competition in science.

\begin{table}
\caption{Example transient facilities and their strengths and
 weaknesses in the area of communicating and coordinating follow-up}
\begin{tabular}{p{6cm}p{6cm}}
\hline
\multicolumn{1}{c}{Strengths} & \multicolumn{1}{c}{Weaknesses} \\
\hline
\\
\multicolumn{2}{c}{Swift}\\
\\
Rapid trigger release & Insufficient optical/IR follow-up
spectroscopic facilities \\ GCN network for alert and prompt data
dissemination & Lack of buy in to VOEvent\\ Motivated follow-up
community & Difficult to coordinate multiple follow-up groups \\
\\
\multicolumn{2}{c}{Palomar Transient Factory and Pan-STARRS-1}\\
\\
Rapid discovery and initial classification & Discoveries mostly
private, most interesting not released \\
Good light curve coverage and spectral follow-up
(different strategies, but both work) & Neither are all sky, still
area limited  \\ 
Large number of interesting sources & Fast prioritisation and
multi-wavelength links could be automated better \\
\\
\multicolumn{2}{c}{Advanced-LIGO and Advanced-VIRGO}\\
\\
New discovery space & Large position errors in early years \\
Willing to work with external teams and release alerts & Need dedicated wide-field follow-up   facilities \\ 
Motivated expert team & Tension between rewarding instrument versus follow-up team \\
\\
\\
\multicolumn{2}{c}{LOFAR}\\
\\
Real-time alerts planned & Delay in implementation of alert system \\
Large simultaneous sky coverage & Small temporal buffer ($\le 1$ minute) \\
Large discovery space & Unknown rapid transient radio sky \\

\hline
\end{tabular}
\end{table}

\section{Future facilities and instrumentation}

As discussed above, there were two science priorities that were
consistently thought to be of such promise that they could potentially
lead to ground breaking new results and potentially impact on
fundamental physics and cosmology.

The first of these was the detection of gravitational waves from
compact binary mergers and short GRB detections (Gehrels 2013, Rosswog
2013). 
 Advanced LIGO's most
promising first sources are likely to be short GRBs from NS-NS or
BH-NS mergers (Rosswog 2013). This discovery requires cooperation between Advanced
LIGO, gamma-ray satellites (of which Swift, Fermi and SVOM are the
most likely), but also a very wide-field optical monitor that can scan
the whole sky every night and efficiently observe large error
regions. The ATLAS project (Tonry 2013) is a pair of 0.5m
telescopes with a 40 square degree camera that can scan the available
sky twice per night to $\sim20^{m}$. It has been given funding from
NASA to enter construction phase. The Next Generation PTF is a project
to fill the Palomar Schmidt focal plane with CCDs to create a 30-40
square degree field-of-view (FoV). Pan-STARRS2 has entered
construction which will give a pair of sensitive 7 square degree FoV
cameras on 1.8m telescopes. These three projects have the potential to
provide the optical follow-up to GW and gamma-ray sources, giving a
powerful combination that could detect GWs and quantify their
sources. This has potential implications far beyond astronomy and will
test fundamental physics if these sources can be found and measured.

The second priority was the discovery and characterisation of high redshift
GRBs. The immense luminosity of GRBs has allowed their detection at
$z\sim9$, but this requires rapid near-infrared follow-up to Swift
triggers and further rapid response from the largest telescopes. The fact that
they can be used as beacons to probe the intervening intergalactic
medium and also locate the first galaxies means that this type of
transient science has exciting potential for early Universe studies. 
This meeting highlighted the importance to have combined facilities of
a gamma-ray burst monitor (Swift and SVOM are the only two missions
currently working or funded), a suite of 4m near-infrared telescopes for rapid
response (and filtering the J+H band drop out afterglows)
and access to 20-40m telescopes (E-ELT, TMT, GMT) at the same
time. While all of these facilities either exist, or are likely to
exist at some time in the future, this group foresees challenges in
them existing at the same time and working coherently together with
large collaborations.

Discussions amongst the teams and in the plenary session also lead to
a conclusion that a wide-field X-ray facility would facilitate both of
these science areas.  A coded-mask instrument operating in the
1-100~keV range with around 5000 sq degrees FoV, combined with a
focusing telescope for softer X-rays (900 sq degrees FoV over ~0.1-10
keV) would target enable rapid and precise localisation of X-ray
sources (targeting soft GRB afterglows) of Advanced LIGO/Advanced
VIRGO gravitational-wave sources and potentially neutrino
transients. This instrument would also be capable of identifying
high-luminosity high-redshift GRBs and provide X-ray detections of
transients located by the new generation of optical wide-field
facilities. The latter include supernova shock break-outs, black hole
tidal disruptions, magnetar flares, and daily X-ray monitoring of
large areas of sky. Such a mission has already been proposed as an
S-class mission to ESA (A-STAR) and its science goals were focused by
the expert gathering at the Kavli Centre.

\section{Acknowledgements}

We are grateful for the generous funding provided by the Royal Society
to enable many participants to attend the meeting and for the
hospitality of their staff at the Kavli Royal Society International
Centre. We thank all of the participants for their enthusiastic
participation and in particular Lars Bildsten, Neil Gehrels and Brian
Schmidt for acting as group leaders. We thank Julian Osborne 
for providing the top panel of Figure 1, and Shri Kulkarni 
for the lower panel of Figure 1.

\end{document}